\documentclass[aps,prd,twocolumn,showpacs,superscriptaddress,groupedaddress,nofootinbib]{
revtex4-1}  
\usepackage{graphicx}  
\usepackage{dcolumn}   
\usepackage{bm}        

\usepackage{amsmath}
\usepackage{amsthm}
\usepackage{amssymb}
\usepackage{amsfonts}
\usepackage{ctable}

\usepackage{bbm}
\usepackage[format=plain,justification=raggedright,textfont=small]{caption}

\definecolor{orange}{RGB}{255,102,0}

\usepackage{ulem}

\begin{document}

\title{Ponderomotive effects in multiphoton pair production}

\author{Christian Kohlf\"urst}
\email{christian.kohlfuerst@uni-jena.de}
\affiliation{Theoretisch-Physikalisches Institut, Abbe Center of Photonics,
Friedrich-Schiller-Universit\"at Jena, Max-Wien-Platz 1, D-07743 Jena, Germany \\ 
Helmholtz-Institut Jena, Fr\"obelstieg 3, D-07743 Jena, Germany}
\author{Reinhard Alkofer}
\email{reinhard.alkofer@uni-graz.at}
\affiliation{Institute of Physics, University of Graz, NAWI Graz, Universit\"atsplatz 5, 8010 Graz, Austria}                        
\date{\today}

\begin{abstract}
The Dirac-Heisenberg-Wigner formalism is employed to investigate electron-positron pair 
production in cylindrically symmetric but otherwise spatially 
inhomogeneous, oscillating electric fields.  The oscillation frequencies are hereby tuned to
obtain multiphoton pair production in the nonperturbative threshold regime.
An effective mass as well as a trajectory-based semi-classical analysis are 
introduced in order to interpret the numerical results for the distribution functions as well as
for the particle yields and spectra. 
The results, including the asymptotic particle spectra, display clear signatures of ponderomotive forces.
\end{abstract}

\maketitle

\section{Introduction}

Electron-positron pair production in strong electric fields,
the Sauter-Schwinger effect, is a 
long-standing theoretical prediction  \cite{Sauter:1931zz} which still
awaits experimental verification. Multiphoton pair production, on the other 
hand, has already been observed in a laboratory \cite{Burke:1997ew}.
The dynamically assisted Sauter-Schwinger effect \cite{PhysRevLett.101.130404,Linder}
exploits the idea that a combination of a low with a high frequency (``multi-photon'') laser pulse
will lead to pair production rates which are significantly larger than the sum of the rates
for the two separate pulses. Together with the arising
capabilities of high-intensity laser technology, 
see, e.g., refs.~\cite{Heinzl:2008an,Marklund:2008gj},
such a combination of laser pulses 
will make experimental tests in this regime of non-linear QED 
in the near future possible.

Besides on the technological progress future experimental tests and their interpretations 
will depend substantially on more reliable calculations which include
the inhomogeneities of the electromagnetic fields as they typically
occur in the focus of crossing laser beams.
Computations for inhomogeneous and time-dependent fields (i.e., 
beyond the previously well-studied case of time-dependent but homogeneous
electric fields) have been started a few years ago and have seen steady 
progress \cite{Dunne:2005sx,Hebenstreit:2011wk,Han:2010rg,Kohlfurst:2015niu}.
These studies are, however, still far from providing a satisfactory
understanding of the effects which originate from the finite spatial extension
of the considered laser pulses. 

As the particle creation rate depends on the laser intensity, see, e.g., 
ref.~\cite{Ringwald:2001ib} for a discussion in the context of the 
planning of the XFEL, one would na\"ively expect 
that better focusing and therefore higher local field intensities lead 
always to an increase of the particle yield.
However, the dynamics in high-intensity laser fields is much more complex 
and results not only in strong but also non-monotonic dependencies of the yield
on the characteristics of the laser field \cite{Linder,Hebenstreit:2011wk,Kohlfurst:2015niu}.
Other quantities like, e.g., distribution functions and particle spectra, display
 a rich structure with sometimes surprisingly large sensitivities to small changes
of the field parameters.

Clearly such a situation calls for the search of concepts which are able to allow
an, at least qualitative, understanding.
One possibility to quantify the effects associated with a spatially inhomogeneous 
field is based on the notion of an effective mass 
which an electron acquires in a background electromagnetic field
\cite{Volkov:1935,mStar:Lasers,Dodin:2008,Kibble:1966zza,Neville:1971,Heinzl,Lavelle:2017dzx}. 
This parameter, as every effective mass, reflects the ``integrated`` 
collective interactions  of a particle with its surroundings.
It thus provides the possibility of a drastic simplification but nevertheless 
allows the coarse-grained description of highly intricate effects.
Although this might be an
oversimplification with respect to details of the resulting spectra, the 
concept of an effective mass
works astonishingly well, a fact which can be attributed 
to the unique conditions in high-intensity laser 
experiments  \cite{mStar:FELs}.
Correspondingly, the idea of an effective mass has been applied recently to 
multiphoton pair production \cite{Kohlfurst:2013ura,Li}.
In these studies, however, the employed electric fields were homogeneous
thereby greatly simplifying the process under investigation. 

In the following we will discuss particle creation in inhomogeneous fields and 
introduce to this end a more general concept for the effective mass and relate it
via a semi-classical analysis to ponderomotive forces.
(NB: A ponderomotive force is a nonlinear force that a ``classical''
charged particle experiences in an inhomogeneous oscillating electromagnetic field, 
{\it cf.} ref.~\cite{Bauer} and references therein.) 
We will concentrate in a first step hereby
on multiphoton pair production  in the nonperturbative threshold regime. 
As is evident from the discussion above, 
in the parameter regions of interest  the  effects of a laser pulse's finite
spatial extension 
cannot be neglected. (For a more detailed discussion of this issue, see,
e.g., ref.\ \cite{Diss}.) Hereby one should distinguish two aspects:
First, how does the finiteness of the fields' extension influence the pair production 
process, and second, how does a spatially inhomogeneous field
alter the subsequent electron/positron dynamics. To keep the calculational complexity in a 
manageable range we will assume cylinder symmetry of the electric field. The spatial dependence
can be inferred, e.g., from a gauge potential which is directed along a direction and is inhomogeneous 
only w.r.t.\ the same direction. This has the advantage that the corresponding magnetic field vanishes.
However, one then clearly has no propagating waves. Nevertheless, such a configuration provides some, 
although simplistic, model of the focus of two counter-propagating lasers.

This paper is organized as follows: In Sect.\ \ref{Multiphoton} several aspects of
multiphoton pair production in the nonperturbative threshold regime 
underlying our work will be briefly summarized:
In subsection \ref{DHW} a short description of the Dirac-Heisenberg-Wigner (DHW) formalism 
for the case of an electric field with cylinder symmetry
is provided; subsection \ref{NumTreat} deals with the numerical treatment via Fourier transform; 
and in subsection \ref{Fields} the form of the electric field employed later on is given.
In sect.\ \ref{Interpretational} two interpretational concepts are introduced and discussed: 
the effective mass and a trajectory-based semi-classical analysis. In sect.\ \ref{Results} numerical 
results, ordered according to the longitudinal and transverse momentum distributions, are presented and
discussed. Sect.\ \ref{Conclusions} contains the conclusions of the presented study. 

Furthermore, we use $\hbar = c = 1$ throughout this paper.

\section{Theoretical description of multiphoton pair production}
\label{Multiphoton}

\subsection{DHW Formalism}
\label{DHW}

The present study is based on the DHW formalism, a relativistic phase-space approach, 
which has been developed for the case of pair production in refs.~\cite{Vasak:1987um}. 
Hereby the electron is treated as a quantum field but the laser pulse is approximated by its mean-field. 
Given the magnitudes of the electric field needed in pair production this is
a justified approximation.

A convenient starting point is provided by the gauge-invariant density operator of the system,
\begin{equation}
 \hat {\mathcal C}_{\alpha \beta} \left( r , s \right) = \mathcal U \left(A,r,s 
\right) \ \left[ \bar \psi_\beta \left( r - s/2 \right), \psi_\alpha \left( r + 
s/2 \right) \right],
\end{equation}
in terms of the electron's spinor-valued Dirac field $\psi_\alpha (x)$.
Hereby $r$ denotes the center-of-mass  and $s$ the relative coordinate. 
The Wilson line factor 
\begin{equation}
 \mathcal U \left(A,r,s \right) = \exp \left( \mathrm{i} \ e \ s \int_{-1/2}^{1/2} d 
\xi \ A \left(r+ \xi s \right)  \right)
\end{equation}
is introduced to render the density operator gauge-invariant. It depends on the elementary charge $e$
and the background gauge field $A$. Treating the background field in Hartree approximation, i.e., 
\begin{equation}
 F^{\mu \nu} \left( {x} \right) \approx \langle \hat F^{\mu \nu} \left( 
{x} \right) \rangle ,
\end{equation}
no path ordering is needed, and in a given Lorentz frame and gauge the background gauge field  
$A \left(\mathbf{x}, t \right)$ is a fixed $c$-number valued function. The
covariant Wigner operator,
\begin{equation}
 \hat{\mathcal W}_{\alpha \beta} \left( r , p \right) = \frac{1}{2} \int d^4 s \ 
\mathrm{e}^{\mathrm{i} ps} \  \hat{\mathcal C}_{\alpha \beta} \left( r , s 
\right),
\end{equation}
reflects thus the quantum fluctuations of the electron but not the one of the laser field.

The main implication of the Hartree approximation for the electromagnetic field
becomes evident when taking the vacuum expectation value of the covariant Wigner operator
to obtain the covariant Wigner function
\begin{equation}
 \mathbbm{W} \left( r,p \right) = \langle \Phi \vert \hat{\mathcal W} \left( r,p 
\right) \vert \Phi \rangle.
\end{equation}
In its equation of motion the electromagnetic field can be factored out, e.g.,
\begin{equation}
 \langle \Phi \vert F_{\mu \nu} \ \hat{\mathcal{C}} \vert \Phi \rangle = 
F_{\mu \nu} \langle \Phi \vert \hat{\mathcal{C}} 
\vert \Phi \rangle
\end{equation}
which allows to resolve the otherwise infinite BBGKY hierarchy of correlation functions. 

Being a Dirac-matrix valued quantity the Wigner function is best decomposed into 16 covariant Wigner 
coefficients 
\begin{equation}
\mathbbm{W} = \frac{1}{4} \left( \mathbbm{1} \mathbbm{S} + \textrm{i} \gamma_5 
\mathbbm{P} + \gamma^{\mu} \mathbbm{V}_{\mu} + \gamma^{\mu} \gamma_5 
\mathbbm{A}_{\mu} + \sigma^{\mu \nu} \mathbbm{T}_{\mu \nu} \right) 
\label{decomp}
\end{equation}
where the corresponding transformation properties are made evident by the notation. 
Working in a definite frame it proves advantageous to 
project on equal times which yields the equal-time Wigner function
\begin{align}
 \mathbbm{w} \left( \mathbf{x}, \mathbf{p}, t \right) = \int \frac{d p_0}{2 \pi} 
\ \mathbbm{W} \left( r,p \right)
\end{align}
and by an analogous decomposition to eq. (\ref{decomp})
the corresponding equal-time Wigner coefficients $\mathbbm{s},\mathbbm{p},
\mathbbm{v}_{0,x,y,z},  \ldots$.

Exploiting cylindrical symmetry, which will be assumed throughout the following,
and keeping the electric field homogeneous in transversal direction results in 
a reduced system of differential equations\footnote{The coefficients given in the following 
are obtained by linear combinations of the equal-time Wigner coefficients. 
The quantity $\overline{\mathbbm{v}}$, for example, is defined as a linear superposition 
of $\mathbbm{v}_z$ and tensor components $\mathbbm{t}_{xz}$
and $\mathbbm{t}_{yz}$.
The details of the derivation can be found in ref. \cite{Diss}.}:
\begin{alignat}{5}
  & D_t \overline{\mathbbm{s}}     && && -2 p_z \overline{\mathbbm{p}} &&+ 2 
p_{\rho} \overline{\mathbbm{v}} &&= 0, \label{eq_Cy1} \\
  & D_t \overline{\mathbbm{v}} &&+\partial_{x} \overline{\mathbbm{v}}_0 && &&- 2 
p_{\rho} \overline{\mathbbm{s}} &&= -2 \overline{\mathbbm{p}},  \\    
  & D_t \overline{\mathbbm{p}} && && +2 p_z \overline{\mathbbm{s}} && &&= 2 
\overline{\mathbbm{v}},  \\ 
  & D_t \overline{\mathbbm{v}}_0 &&+ \partial_{z} \overline{\mathbbm{v}} && && 
&&= 0, \label{eq_Cy4}  
\end{alignat}
where the pseudo-differential operator $D_t$ reads
\begin{align}
 D_t = \partial_{t} + e \int \ d\xi \ E \left(z+\textrm{i} \xi \partial_{p_z},t 
\right) \ \partial_{p_z}. \label{Pseudo}
\end{align}
We refrained from putting a spatial index on the electric field $E$ as by construction  
it is oriented in the $z$-direction.

The advantage of a representation with Wigner coefficients lies in the 
fact, that it allows to identify $\overline{\mathbbm s}$ as mass, 
$\overline{\mathbbm v}_0$ as charge
and $\overline{\mathbbm v}$ as current density \cite{Vasak:1987um}.
In order to perform calculations within the DHW formalism
we employ vacuum initial conditions: 
\begin{align}
  \overline{\mathbbm{s}}_{in} = -\frac{2}{\omega},\ \overline{\mathbbm{v}}_{in} = 
-\frac{2 p_z}{\omega} ,\ \overline{\mathbbm{p}}_{in} = -\frac{2 p_{\rho} }{\omega},
\end{align}
where $\omega = \omega \left(p_z,p_{\rho} \right) =  \sqrt{m^2 + p_z^2 + p_{\rho}^2}$
is the one-particle energy. It is convenient to subtract these initial conditions,
and therefore we define
\begin{equation}
 \overline{\mathbbm{w}}^v := \overline{\mathbbm{w}} - \overline{\mathbbm{w}}_{in},
\end{equation}
where $\overline{\mathbbm{w}}$ is a placeholder for any Wigner component.

Additionally, we
define the particle number density per unit volume in momentum space
\begin{align}
 N \left(p_z, p_{\rho} \right) = & \ \int dz \ \frac{
\overline{\mathbbm{s}}^v + p_z  \overline{\mathbbm{v}}^v + p_{\rho} \overline{\mathbbm{p}}^v }
{\omega \left(p_z,p_{\rho} \right)} \, .\label{NN}
\end{align}
Consequently, the total particle yield per unit volume is defined via
\begin{align}
 N = & \int dp_z \ \int dp_{\rho} \ N \left(p_z, p_{\rho} \right). 
\label{NNfull}
\end{align}

\subsection{Fourier transform and numerical treatment}
\label{NumTreat}

Eqs.\ \eqref{eq_Cy1}-\eqref{eq_Cy4} can be solved numerically  
without any further approximations or truncations \cite{Hebenstreit:2011wk,Diss}.
The challenging part is the non-locality of the pseudo-differential 
operator \eqref{Pseudo} which can be nevertheless treated 
if the electric field $E \left( z, t \right)$ 
can be Taylor expanded and integrated sufficiently efficient.
The differential operator \eqref{Pseudo} splits naturally in two parts:
\begin{equation}
 D_t = \partial_{t} + e \int d\xi \ E \left( z+\textrm{i} \xi \partial_{p_z},t 
\right) \partial_{p_z} =: \partial_{t} + \Delta. \label{Pseudo2}
\end{equation}
To apply the operator $\Delta$ on the (subtracted) Wigner components
we Fourier transform and inverse Fourier transform 
$\Delta  \overline{\mathbbm{w}}^v$
w.r.t.\ to the variable $p_z$,
i.e., we employ 
\begin{equation}
f \left( p_z \right) = 
 {\cal F}_{p_z}^{-1} \left[ {\cal F}_{p_z}[ f \left( p_z \right)] \right] = 
{\cal F}_{p_z}^{-1} \left[ \tilde f \left( {\rm k}_{p_z} \right)\right] ,
\end{equation}
Taylor expand the electric field, use   
\begin{equation}
 {\cal F}_{p_z}[\frac{d^n}{dp_z^n}f(p_z)]= \left( \textrm{i} {\rm k}_{p_z} \right)^n \tilde f \left( 
{\rm k}_{p_z} \right) ,
\end{equation}
and resum then to obtain the generic form (for more details see ref.~\cite{Diss}):
\begin{align}
 &\Delta \ \overline{\mathbbm{w}}^v \left( z, p_z, t \right) \notag \\
 &= \mathcal {F}_{p_z}^{-1} [ \textrm{ie} {\rm k}_{p_z} \int d\xi \ E \left( z-\xi 
{\rm k}_{p_z},t \right) \ \tilde{\overline{\mathbbm{w}}} \left( z, {\rm k}_{p_z}, t \right) 
].
\end{align}

Due to the fact, that a Fourier transform takes into account all
points in a domain, the introduction of global basis functions turned out to be 
favorable compared to the finite-difference-method used in ref.~\cite{Hebenstreit:2011wk}.
Hence, we have equidistantly discretized spatial and momentum directions, 
respectively, turning the system of partial differential equations 
\eqref{eq_Cy1}-\eqref{eq_Cy4} into
a high-dimensional ($N_z \times N_{p_\rho} \times N_{p_z}$) system of ordinary 
differential equations with $t$ as the only continuous parameter. 

Furthermore, we choose periodic boundary conditions in $z$ and $p_z$,
\begin{equation}
 \overline{\mathbbm{w}}^v \left( z_0,p_z,t \right) = \overline{\mathbbm{w}}^v \left( 
z_{N_z},p_z,t \right) 
 \end{equation}
 and
 \begin{equation}
 \overline{\mathbbm{w}}^v \left( z, p_{z,0},t \right) = \overline{\mathbbm{w}}^v 
\left(z,p_{z,N_{p_z}},t \right),
\end{equation}
respectively.
Additionally, we set
\begin{align}
  \overline{\mathbbm{w}}^v \left( z_{k_i}, p_{z,{k_j}},t \right) = 0 \quad 
\textrm{if} \quad
 k_i = 0 \ \textrm{or} \ k_j = 0.
\end{align}
These choices do not influence the numerical results as long as 
the chosen discretized domain is sufficiently large, 
and the number of grid points in every direction is sufficiently high. 

After discretizing  
 eqs.\ \eqref{eq_Cy1}-\eqref{eq_Cy4} in this way
they can be solved using pseudospectral methods \cite{Boyd}. 
The time integration is done using a Dormand-Prince Runge-Kutta integrator of 
order 8(5,3) \cite{NR}.

\subsection{Model for the fields}
\label{Fields}
 
As stated above, if the spatial dependence of the electric field 
is inferred from a gauge potential which is directed along a direction and is inhomogeneous 
only w.r.t.\ the same direction the corresponding magnetic field vanishes. In addition, 
two time scales are needed to tune a pulse of finite duration to the multiphoton regime. These
requirements are fulfilled by the ansatz
\begin{align}
 E \left( z,t \right) &= \varepsilon E_0 \ {\cal E} \left( z \right) \ F \left( t \right) 
\notag \\
 &= \varepsilon E_0 \ {\cal E} \left( z \right) \ \cos^4 \left( \frac{t}{\tau} \right) \ 
\cos \left( \omega t \right), \label{EE0}
\end{align}
for $t \in \left[ -\pi \tau/2,~ \pi \tau/2 \right]$, and $E=0$ otherwise. 
Hereby the critical field strength $E_0=m^2/e$ has been factorized out for convenience. 
Non-perturbative multiphoton pair production is probed if one choses
the product $\omega \tau > 1$ and a Keldysh parameter of $\gamma = \omega / m \varepsilon > 1$ \cite{Diss}. 

As we want to investigate how focusing influences the particle distribution rate we chose a well-localized
electric field with a Gaussian shape of width $\lambda$: 
\begin{equation}
 {\cal E} \left( z \right) = \exp \left( -\frac{z^2}{2 \lambda^2} \right). 
\label{EE1}
\end{equation}

\section{Semi-classical analysis}
\label{Interpretational}

In Sect.\ \ref{Results} it will become obvious that the dependencies of observables
on the field parameters are of an astonishing complexity. In order to obtain an interpretation and
such an understanding of the results we will analyze them referring to the concepts developed in this
section.

Introducing a generalized effective mass concept and relating it to arising 
ponderomotive forces is a central aspect of this paper. 
We discuss the improvements compared to previous definitions of the effective 
mass in the context of pair production in section \ref{Ch_Con1}. Moreover,
based upon refs. \cite{Bauer,Pond}, we draw a connection between a spatially 
dependent effective mass and ponderomotive forces. 
However, 
an interpretation of the final particle distribution on the basis of an 
effective mass becomes more involved for spatially inhomogeneous fields due to 
the position-dependence of the gradient. 
Hence, we rely on a semi-classical trajectory-based model, which allows us to determine 
the overall scheme by simple means. 

\subsection{Effective mass and ponderomotive forces}
\label{Ch_Con1}

Various studies have used effective masses to simplify intermediate calculations 
\cite{Kibble:1966zza,Neville:1971,Mackenroth:2010jr,Kibble}.
It was suggested only recently to employ the concept of an effective mass to 
determine directly observable quantities regarding particle creation, see ref. 
\cite{Kohlfurst:2013ura}.  

In case of a monochromatic plane wave, the effective mass takes the form 
\cite{Volkov:1935}:
\begin{equation}
 M_{\ast} = m \sqrt{1 + \xi^2} ,~ \textrm{with} ~ \xi = \frac{e}{m} \sqrt{- 
\langle A^{\mu} A_{\mu} \rangle }. 
\end{equation}
More general definitions have been proposed in ref. \cite{Kibble}.
We, however, adopt essentially but modify slightly the definition above 
and parameterize the effective mass as follows
\begin{align}
 m_{\ast} \left( \mathbf{x} \right) &= m \sqrt{1+\tilde{\xi} \left( \mathbf{x} \right)^2}, \\ 
   &\tilde{\xi} \left( \mathbf{x} \right) = \frac{e}{m} \sqrt{- \langle A_{\mu} \left( 
\mathbf{x},t \right) A^{\mu} \left( \mathbf{x},t \right) \rangle }. \label{meff}
\end{align}
Similarly to refs. \cite{Kohlfurst:2013ura,Kibble}, we cope with the temporal 
finiteness of the pulse by averaging over one field oscillation around $t=0$ 
only; this approximation is well justified for long, flat-topped multicycle pulses due
to the minor influence of the envelope function.

The relativistic ponderomotive force then yields, see ref. 
\cite{Bauer},
\begin{equation}
 F_{p} = -\left( \mathbf{v}_0 \cdot \boldsymbol{\nabla}_x m_{\ast},~ 
\boldsymbol{\nabla}_x m_{\ast} + \frac{\gamma_0 -1}{v_0^2} \left( \mathbf{v}_0 
\cdot \boldsymbol{\nabla}_x m_{\ast} \right) \mathbf{v}_0 \right), 
\label{Pond_Rel}
\end{equation}
where $\gamma_0$ is the Lorentz factor and $\mathbf{v}_0$ denotes the velocity 
of the quasi-particle. 

We do not use equation \eqref{Pond_Rel} directly to calculate reference values.
However, as equation \eqref{Pond_Rel} describes an effective force, where all short scale contributions are ''integrated out``,
it primarily serves as a tool helping to interpret the results obtained from solving the system \eqref{eq_Cy1}-\eqref{eq_Cy4}.
Analyzing equation \eqref{Pond_Rel} analytically, we can deduce that the term $\boldsymbol{\nabla}_x m_{\ast}$ is
the decisive factor in order to understand the effective force on the particles.
In turn, we expect that all particles 
are forced from strong-field towards weak-field regions.
Furthermore, in the case under consideration we 
primarily expect a boost in the parallel momentum.
If particles are created with vanishing transversal momentum, ponderomotive 
forces only act upon them in direction parallel to the applied field. 

\subsection{Trajectory-based semi-classical model}
\label{Ch_Con2}

The virtue of the DHW method, namely to take into account various effects
in one common approach, obstructs the analysis of its results.
Hence, we introduce a trajectory-based model in order to overcome these difficulties. 

Contrary to Schwinger pair production the formation time in multiphoton pair 
production is quite long. For the sake of simplicity, we still want to assume
that particles are created at points in space-time, where the electric field 
takes on its maximal values.\footnote{We are well aware of the fact that this assumption 
leads to an oversimplification of the 
description of multiphoton processes as particles can be created at all times.
However, evaluating particle trajectories that start at turning points of the
applied fields yields reference values for the final particle momenta which suffices, 
in the context of the current investigation, to validate the method.}
Moreover, for the case of 
$n$-photon pair production we expect that the initial 
particle momenta $p$ can be derived via energy conservation laws 
\cite{Delone:1994,Kohlfurst:2013ura}:
\begin{equation}
 p_n^2 = \left( \frac{n \omega}{2} \right)^2 - m_{\ast}^2. \label{En}
\end{equation}
These assumptions are sufficient 
to determine the trajectory of
a particle in an external  field as they provide all 
needed initial conditions. 

Due to the form of the electric field and especially due to the absence of a 
magnetic field, spin-effects 
will be ignored at this point. 
Hence, we employ the relativistic Lorentz equation 
\begin{equation}
 \frac{d P^{\mu}}{d \tau} = e F^{\mu \nu} U_{\nu} \label{Lorentz}
\end{equation}
to analyze the particle's trajectory in the external field. 
Here, $F^{\mu \nu}$ is the electromagnetic field strength tensor, $U_{\nu}$ the 
four-velocity and $P^{\mu}$ the four-momentum. 

This method is a convenient yet powerful tool to analyse
the dynamics of pair production. 
However, it should be understood as a simple approximation, which clearly cannot 
replace a full quantum field theoretical treatment of the process. 
In ref. \cite{Diss} its usefulness is demonstrated for a variety of field 
configurations. 

\section{Numerical results for multiphoton pair production in the threshold regime}
\label{Results}

In the following we discuss the solutions of the PDEs 
\eqref{eq_Cy1}-\eqref{eq_Cy4} and compare the outcome with the results obtained 
from the trajectory-based approach.
We analyze the distribution function for
parallel particle momenta and subsequently the total production rate. 

For the pulse length we have chosen a value of $\tau = 100$ $m^{-1}$ for all 
calculations. Such a pulse length is sufficient to capture all essential features of 
multiphoton pair production. 



\subsection{Parallel momentum distribution}
\label{Ch_Res1}

\begin{figure}[th]
\begin{center}
\includegraphics[width=0.48\textwidth]{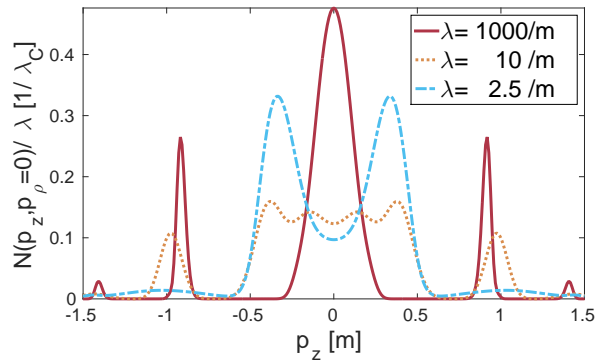}
 \end{center}
 \caption{Reduced particle distribution function obtained from a DHW calculation for a field of 
strength $\varepsilon=0.5$, length $\tau=100$ $m^{-1}$ and frequency 
$\omega=0.7$ $m$. The peaks can be related to $n$-photon pair production. The 
stronger the focus the stronger the ponderomotive forces, thus the higher the 
final particle momentum.} 
 \label{Fig1}
\end{figure}

In this subsection we only analyze particle spectra where $p_\rho=0$.
In Fig. \ref{Fig1} a typical spectrum for multiphoton-dominated particle creation is displayed. 
Especially for quasi-homogeneous fields, $\lambda=1000$ $m^{-1}$, the characteristic multiphoton 
peaks are easily distinguishable.
Here, the peak in the middle stems from a \mbox{$3$-} photon process and the 
side maxima are related to \mbox{$4$-}, and $5$-photon pair production.


However, comparing the relative peak sizes with results obtained from calculations within
homogeneous fields one finds that the strong peak around $p_z=0$ is 
now more pronounced.
Nevertheless, we recover the quantum kinetic limit \cite{Smolyansky:1997fc} when 
evaluating the results for $\lambda \to \infty$ at $z=0$.

Going beyond this limit reveals that
a decrease of the pulse's spot size leads to dramatic changes in the distribution function. 
In case of, e.g. $\lambda=10$ $m^{-1}$, the dominant peak in the particle momentum spectrum 
takes on a much wider form
which can be related to quantum interferences. 
It is known from atomic ionization \cite{Corkum} that $n$-photon peaks can be 
interpreted as a result of particle trajectories adding up to the interference 
pattern observable. 
Around $\lambda = 10$ $m^{-1}$ the finite size of the laser pulse seems to prevent a coherent 
superposition. In turn, the corresponding interference pattern become disturbed 
resulting in the broadened distribution. 
A quantitative comparison between the trajectory-based approach and the DHW 
formalism is summarized in Tab. \ref{Tab1}.

At an even smaller focus size, $\lambda = 2.5$ $m^{-1}$, we observe so-called peak splitting.
This effect can be understood in the context of ponderomotive forces.
Concentrating on the $3-$photon peak, particles are created with close to vanishing momentum $p_z$
and subsequently follow the oscillations of the background field. 
Due to the spatial inhomogeneity these particles then drift to low-intensity 
regions in space. As the applied electric field is symmetric in $z$ there are, however, 
two equally likely options: either the particles are accelerated in $z$ or in $-z$ direction.
Hence, the two peaks at nonzero final momentum. 

\setlength{\extrarowheight}{3pt}
\setlength\tabcolsep{15pt}
\begin{table}[bh]
\caption{Trajectory analysis of particles within an 
external field. The results are obtained by evaluating the relativistic Lorentz 
force equation 
for particles seeded at $t_0=0,~ z_0=0$ in a field of strength $\varepsilon=0.5$, length $\tau=100$ $m^{-1}$, frequency $\omega=0.7$ $m$ and 
spatial extent $\lambda$. 
The different initial momenta $p_{z,0}$ correspond to different $n$-photon 
processes. The final momenta $p_{z,f}$ are obtained at asymptotic times. For 
comparison we provide the results from
a DHW calculation $p_{\rm DHW}$.}
\label{Tab1}
\begin{ruledtabular}
\begin{tabular}{cccc}
    $\lambda$ $[m^{-1}]$ & $p_{z,0}$ $[m]$ & $p_{z,f}$ $[m]$ & $p_{\rm DHW}$ $[m]$  \\
    \colrule
    1000 & 0 & $10^{-7}$ & 0 \\  
    10 & 0 & 0.162 & 0 - 0.38 \\ 
    2.5 & 0 & 0.444 & 0.33 \\    
    \colrule
    1000 & 0.92 & 0.92 & 0.92 \\  
    10 & 0.92 & 0.99 & 0.98 \\ 
    2.5 & 0.92 & 1.12 & 1.03 \\    
    \colrule
    1000 & 1.4 & 1.4 & 1.40 \\  
    10 & 1.4 & 1.43 & 1.43 \\ 
    2.5 & 1.4 & 1.65 & 1.45 \\   
\end{tabular}
\end{ruledtabular}
\end{table}
\setlength{\extrarowheight}{0pt}
\setlength\tabcolsep{0pt}

Particles with non-zero initial momenta acquire only an additional push due to 
ponderomotive forces. 
The strength of this boost depends on the initial conditions, but seems to be 
non-monotonic following the results in Tab. \ref{Tab1}. 
We interpret the increase in momentum as a consequence of particle acceleration 
out of the strong field region within one half-cycle. 
If the spatial extent of the background field is sufficiently small these 
particles are basically unaffected by the following field oscillations and keep 
their respective momentum. 

\begin{figure}[th]
\begin{center}
 \includegraphics[width=0.48\textwidth]{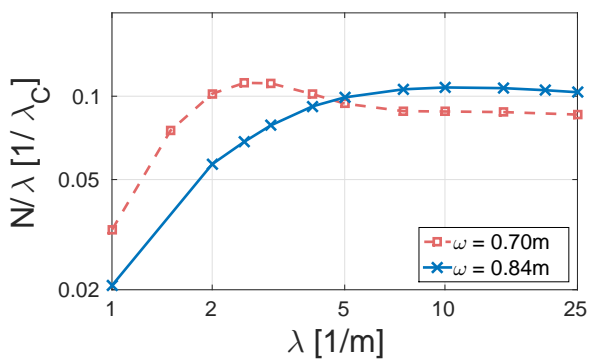}
 \end{center}
 \caption{Log-log plot of the reduced particle yield $N/\lambda$
calculated from the DHW equations for various 
field frequencies $\omega$ as a function of the spatial extent $\lambda$.
The sharp decrease for $\lambda \sim \mathcal{O} \left( 1 \right)$ is directly related to the
lack of sufficient field energy to produce particles. 
Parameters: $\varepsilon = 0.5$, $\tau=100$ $m^{-1}$.} 
 \label{Fig4}
\end{figure}

{A finite spatial extent also affects the particle production rate. 
Fig. \ref{Fig4} shows the reduced particle yield for different photon 
frequencies and various spot sizes. In the case of $3$-photon pair production (blue line in Fig. \ref{Fig4}) the 
introduction of a spatial extent only lowers the particle production rate. 
However, in case of photon frequencies of $\omega = 0.7$ $m$ and a field 
strength of $\varepsilon=0.5$ a $3$-photon pair production process is 
not possible due to energy conservation laws, see equation \eqref{En} and ref. \cite{Kohlfurst:2013ura}.
The fact, that we still observe a peak at $p_z \sim 0$ in the particle spectrum, see Fig. \ref{Fig1},
is a hint towards a dynamically assisted tunneling process, where absorbing $3$ photons
lowers the energy barrier and, in turn, increases the likelihood of a tunneling process to happen.
Additionally, the $4$-photon creation process contributes towards the yield
in Fig. \ref{Fig4} (dashed red line).
All in all, it seems as if especially particles created via the assistance mechanism 
\cite{Linder} benefit from a small spot size and the resulting strong ponderomotive forces leading to an 
overall increase in the reduced yield. Concerning the drop-off for small values of the spatial extent 
$\lambda$: If the spatial extension of the field is of the order of the Compton wavelength the total 
electric field energy 
becomes too small to produce a sizable amount of particles \cite{Hebenstreit:2011wk}.}


\subsection{Transversal momentum distribution}
\label{Ch_Res2}

Following the discussion on particle acceleration parallel to the applied field we now turn our 
attention towards electrons/positrons, which are created with non-vanishing 
transversal momentum. For demonstration purpose we have chosen a slightly higher 
field frequency. In this way, the implications of ponderomotive forces in the 
particle momentum spectrum become more evident. 
The Keldysh parameter for the field used is $\gamma = 1.68$ indicating a process 
in the crossover regime with multiphoton dominance \cite{Diss}. 

\begin{figure}[bh]
 \includegraphics[width=0.49\textwidth]{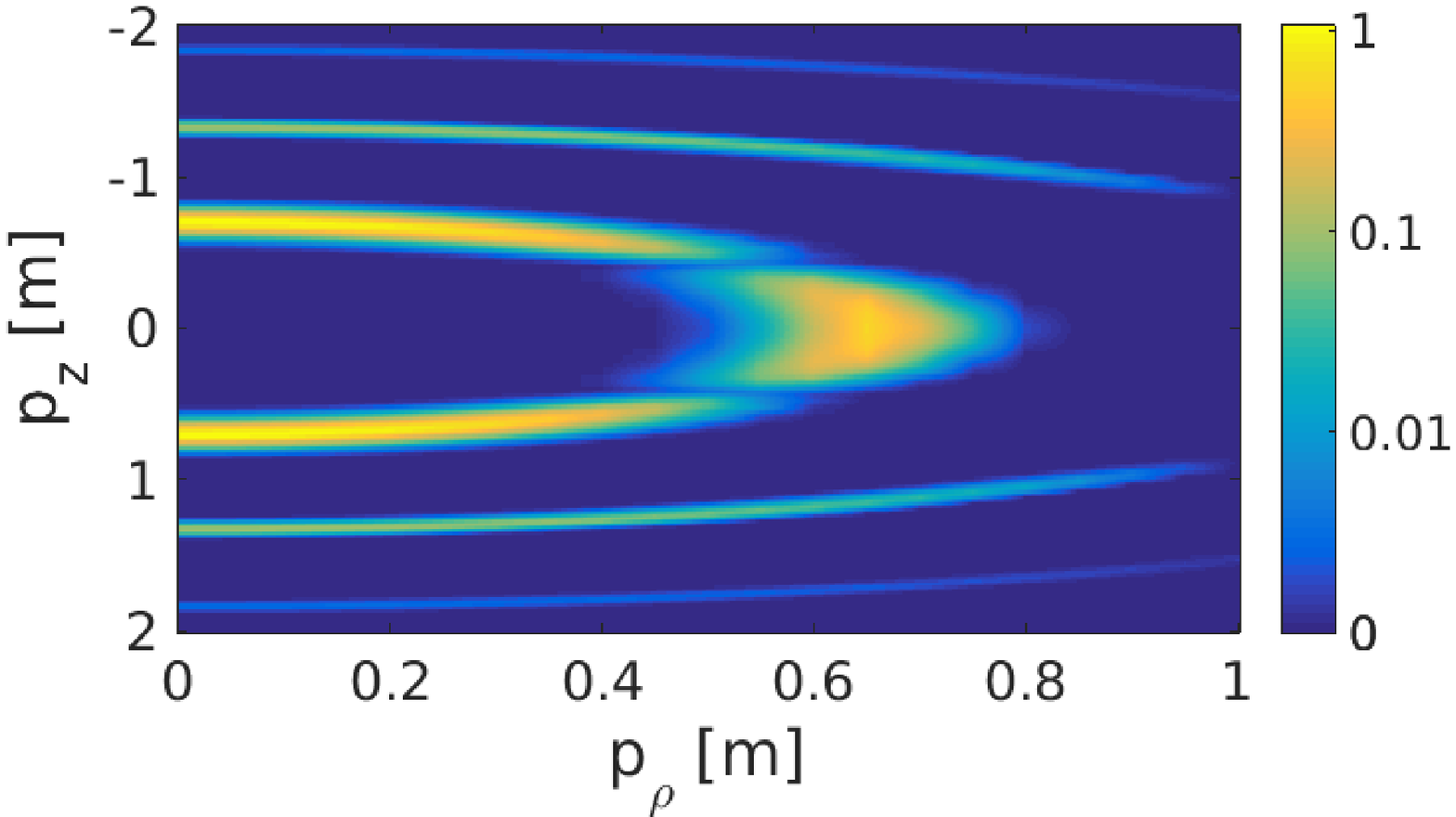}
 \includegraphics[width=0.49\textwidth]{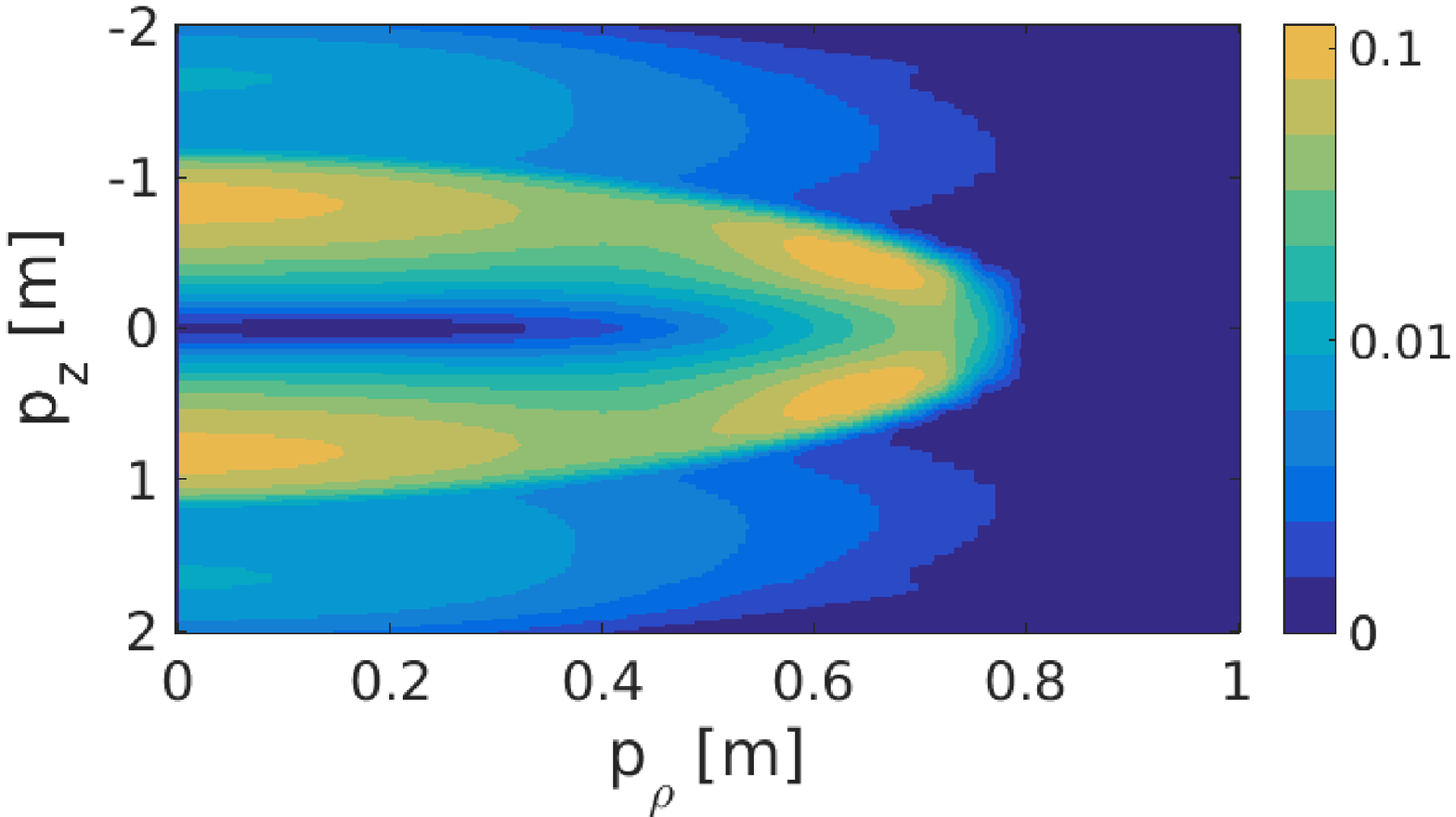}
 \caption{Momentum maps $N/\lambda$ obtained from evaluating the DHW
 equations. The electric field features a field strength of $\varepsilon=0.5$, a pulse length 
of $\tau=100$ $m^{-1}$, a field frequency of $\omega=0.84$ $m$ and spatial extents
$\lambda=100$ $m^{-1}$(top) or $\lambda=2$ $m^{-1}$(bottom). The additional acceleration 
in parallel direction (in the second plot) can be attributed to strong ponderomotive forces. The result
is line broadening and peak splitting.}
 \label{Fig2}
\end{figure}

The comparison between a flat ($\lambda = 100$ $m^{-1}$) and a sharp peak 
($\lambda = 2$ $m^{-1}$) is illustrated in Fig. \ref{Fig2} with focus on a $3$-photon 
process. 
The figure at the top shows a situation close to a result obtained via a purely 
time-dependent electric field \cite{Paulus,Li}. 
Particles created via photon absorption obtain a well-defined total momentum 
\eqref{En}. Additionally, the resulting ring-like structure is superimposed by 
quantum interferences \cite{Delone:1994,Diss}. 

\setlength{\extrarowheight}{3pt}
\setlength\tabcolsep{15pt}
\begin{table}
\caption{
The relativistic Lorentz force equation is evaluated for particles seeded at 
local maxima of the field, $t_0$ and $z_0=0$, in a field of strength $\varepsilon=0.5$, length $\tau=100$ $m^{-1}$, frequency $\omega=0.84$ $m$ and spatial 
extent $\lambda$. 
The initial momenta were chosen to be $p_{z,0} = 0.665$ $m$ and $p_{\rho,0}=0$ 
resembling the photon peak positions in Fig. \ref{Fig2}.}
\label{Tab4}
\begin{ruledtabular}
\begin{tabular}{rcc}
    $t_0$ $[m^{-1}]$ & $p_{z,f} 
\left( \lambda=100 \right)$ $[m]$ & $p_{z,f} 
\left( \lambda=2 \right)$ $[m]$ \\
    \colrule
    $-2 \pi / \omega$ & 0.70 & 0.79 \\  
    $-\pi / \omega$ & 0.70 & 0.85 \\ 
    $0$ & 0.70 & 0.80 \\ 
    $\pi / \omega$ & 0.70 & 0.84 \\    
    $2 \pi / \omega$ & 0.70 & 0.80 \\         
\end{tabular}
\end{ruledtabular}
\end{table}
\setlength{\extrarowheight}{0pt}
\setlength\tabcolsep{0pt}

A smaller spatial extension of the laser focus leads to an increase in strength of 
ponderomotive forces, clearly visible in Fig. \ref{Fig2} in the form of an acceleration in $p_z$-direction. 
In similarity to the previous case, the bunch of particles at
$p_z = 0$, $p_\rho = 0.7$ $m$ is boosted in either $p_z$-direction leading to 
peak splitting. Moreover, the interference pattern diminishes and the sharp 
peaks signalizing $n$-photon processes are washed out. Once again, we analyze
the particle trajectories to understand this line broadening. 
Constraining ourselves to the center of the laser pulse we seed the particles at local maxima of the
applied electric field and calculate the particles final momenta in dependence of the spatial extent, 
see Tab. \ref{Tab4}. In the case of a broad 
spatial extent all particles acquire the same final momenta. 
In case of a small extent, however, the finiteness of the pulse plays a crucial role 
because then different particle creation times lead to different behaviour.

\setlength{\extrarowheight}{3pt}
\setlength\tabcolsep{15pt}
\begin{table}[bh]
\caption{Evaluating the relativistic Lorentz force equation for particles seeded 
at $t_0=0,~ z_0=0$ with parallel momentum $p_{z,0}=0$ and transversal momentum 
$p_{\rho,0}$. The external field shows a strength of $\varepsilon=0.5$, a pulse
length of $\tau=100$ $m^{-1}$ and a frequency of $\omega=0.7$ $m$. The spatial 
extent $\lambda$ is varied. The final parallel momenta $p_{z,f}$ are obtained at 
asymptotic times. Note, that the transversal momenta are nearly unaffected by the external field.}
\label{Tab2}
\begin{ruledtabular}
\begin{tabular}{ccc}
    $p_{\rho,0}$ $[m]$ $\sim$ $p_{\rho,f}$ $[m]$ & $\lambda$ $[m^{-1}]$ & $p_{z,f}$ $[m]$ \\
    \colrule 
    0 & 1000 & $8 \cdot 10^{-6}$ \\   
    0.5 & 1000 & $7 \cdot 10^{-6}$ \\ 
    1 & 1000 & $4 \cdot 10^{-6}$ \\  
    \colrule
    0 & 20 & 0.026 \\   
    0.5 & 20 & 0.021 \\
    1 & 20 & 0.013 \\    
    \colrule
    0 & 10 & 0.16 \\   
    0.5 & 10 & 0.12 \\  
    1 & 10 & 0.07 \\  
\end{tabular}
\end{ruledtabular}
\end{table}
\setlength{\extrarowheight}{0pt}
\setlength\tabcolsep{0pt}

Due to the fact, that we seed
the particles at maxima of the field only, we obtain an upper and a lower limit for the particles' final momenta.
Assuming that electrons and positrons are created also at intermediate times, 
we expect that their respective final momenta
are distributed in between these two limits. It is therefore not surprising that the
peaks in the momentum spectrum appear much wider compared to the quasi-homogeneous case.

The smaller boost for higher transversal momenta 
is a consequence of 
relativistic mechanics as can be seen by comparison with the effective ponderomotive forces \eqref{Pond_Rel}. 
For the sake of completeness, we provide the data obtained through a trajectory analysis in Tab. 
\ref{Tab2}. 

\section{Conclusions}
\label{Conclusions}

We have presented numerical solutions describing multiphoton pair production for oscillating, 
spatially inhomogeneous electric fields in the DHW formalism.
Spatial inhomogeneities introduce effective ponderomotive forces, which 
directly affect the particle momentum distribution and subsequently the total 
yield. Moreover, we have shown that these forces can be understood via a generalized effective mass concept. 

With the aid of a semi-classical trajectory-based model we produced reference values to 
analyse our findings regarding
new phenomena connected with a finite spatial pulse size: Peak splitting and line broadening. 
As for the first effect, we note that the peaks split due to strong  ponderomotive forces 
altering the particle momentum spectrum. Line broadening happens because the particle spectra properties 
which are characteristic for multiphoton pair production erode and instead of sharp lines one obtains
broad bunches.

In summary, we presented here further evidence how important it is to take spatial inhomogeneities of the
fields underlying pair production processes into account.
Therefore, further investigations aiming at an understanding of matter creation from fields
will have, at least, to include spatial variations of the fields.
Given sufficient computational resources the DHW formalism can be readily extended 
such that inhomogeneous magnetic fields can be included.
This would imply a major step towards understanding multiphoton pair production in realistic scenarios further
closing the gap between theory and experiment.

\section{Acknowledgements}
We thank T. Heinzl for helpful and interesting discussions.
C. K. acknowledges financial support by the FWF (Austrian Science Fund)  
Doctoral Program DK W1203-N16, 
by the BMBF under grant No.\  05P15SJFAA (FAIR-APPA-SPARC), and by the
Helmholtz Association through the Helmholtz Postdoc Programme (PD-316). 
Computations were performed on the ``Supermicro Server 1028TR-TF'' in Jena, which
was funded by the Helmholtz Postdoc Programme (PD-316).




\end{document}